\documentclass[10pt,letterpaper,twocolumn]{article} 

\usepackage{ol2}
\usepackage[draft]{hyperref}
\usepackage{amsmath}

\begin{document}

\twocolumn[ 

\title{Nonlocal bright spatial solitons in defocusing Kerr media \\
supported by PT symmetric potentials}
\author{Zhiwei Shi,$^{1}$ Huagang Li,$^{2}$ Xing Zhu,$^3$ and Xiujuan Jiang$^{1,*}$}
\address{$^1$School of Information Engineering, Guangdong University of Technology, Guangzhou 510006, P.R.China
\\
$^2$Department of Physics, Guangdong university of Education, Guangzhou 510303, P.R.China
\\
$^3$State Key Laboratory of Optoelectronic Materials and Technologies, Sun Yat-Sen University,
 Guangzhou 510275, P.R.China
\\
$^*$Corresponding author: jiangxj@gdut.edu.cn}

\begin{abstract}It is studied the bright spatial solitons in nonlocal defocusing Kerr media with parity-time (PT) symmetric potentials. We find that these solitons can exist and be stable over a different range of potential parameters. The influence of the degree of nonlocality on the solitons and the transverse energy flow within the stable solitons are also examined.
\end{abstract}


 ] 

\noindent Generally, defocusing Kerr nonlinearity can induce an enhanced beam
broadening so that it does not support any localized structures other than vortex and dark solitons, which require
background beams~\cite{1,2}. However, the bright solitons, which were thought to exist
only for focusing nonlinearity~\cite{1}, can in fact exist in periodic photonic lattices or waveguide arrays with defocusing
nonlinearity~\cite{3,4,5}. This means that the strong transverse periodic refractive index potentials can not only suppress the beam broadening due to diffraction but also overcome the broadening effect due to the defocusing Kerr nonlinearity~\cite{1}.

The nonlocal nonlinear behavior in a system is related to the change of  nonlinear refractive index, which depends not only on the intensity of a local beam but also on the intensity
of the surrounding region due to a specific distribution
~\cite{5-1}. The nature and
extent of nonlocality substantially depend on the materials. Principally, new effects attributed to nonlocality
have been studied in thermo-optic media~\cite{5-2}, photorefractives~\cite{5-3}, and liquid crystals~\cite{5-5,5-6}.

The definition of PT potentials and their properties were discussed in the past few years~\cite{18,20,21}. The real
part of PT symmetric potentials must be a symmetric
function of position, while the imaginary component should
be antisymmetric. Recently, parity-time symmetric potentials have been introduced into optical field~\cite{6,7,8,9,10,11,12,13,14,15,16,16-1,16-2,16-3}. However, thus far all studies focus on bright solitons in self-focusing optical
PT symmetric media, and bright spatial solitons in nonlocal Kerr self-defocusing  media with a single PT complex potential are never reported.

In this paper, we investigate the bright spatial solitons in nonlocal defocusing Kerr media with parity-time (PT) symmetric potentials.
It is found that these self-tapped states can exist and be stable over a different range of potential parameters.
In addition, we show the influence of the degree of nonlocality on the solitons and the transverse energy flow within the stable solitons.

In a nonlocal Kerr self-defocusing medium with PT symmetric potentials, the one-dimensional optical beam evolution is governed
by the following normalized nonlinear Schr\"odinger-like equation for $q$ and $\phi$, which are respectively the dimensionless light field amplitude and the nonlinear correction to the
refractive index~\cite{6,7,8,9,10,11,12,13,14,15,16,16-1},
\begin{subequations}
\label{eq:one}
\begin{equation}
i\frac{\partial q}{\partial z}+\frac{\partial^2 q}{\partial x^2 }+[V(x)+iW(x)]q-\phi q=0,\label{1a}
\end{equation}
\begin{equation}
\alpha^{2}\frac{\partial^{2}\phi}{\partial x^{2}}-\phi+|q|^{2}=0.\label{1b}
\end{equation}
\end{subequations}
where $z$ is the propagation distance, $V(x)$ and $W(x)$ are the real and the imaginary components of the complex
PT symmetric potential, respectively. $V(x)$ is an even function and $W(x)$ is odd. Physically, $V(x)$ is associated with index
guiding while $W(x)$ represents the gain/loss distribution of the optical potential.
We assume that the depth of the refractive index modulation
is small compared to the unperturbed index. The nonlocality of the materials is supposed to be ruled with an exponential response function $g(x)=1/(2\alpha^{1/2})\exp(-|x|/\alpha^{1/2})$ (as in liquid crystals), where $\alpha$ is the degree of the nonlocality. We are going to search for a stationary soliton solution of Eq.~(\ref{eq:one}) in the form of $q(x,z)=u(x)e^{ibz}$, where $u$ is a complex function and $b$ is the propagation constant of spatial solitons. In this case $u$ satisfies
\begin{subequations}
\label{eq:two}
\begin{equation}
i\frac{\partial u}{\partial z }+\frac{\partial^2 u}{\partial x^2 }+[V(x)+iW(x)]u-\phi u=0,\label{2a}
\end{equation}
\begin{equation}
\alpha^{2}\frac{\partial^{2}\phi}{\partial x^{2}}-\phi+|u|^{2}=0.\label{2b}
\end{equation}
\end{subequations}

\begin{figure}[htb]
\centerline{
\includegraphics[width=8cm]{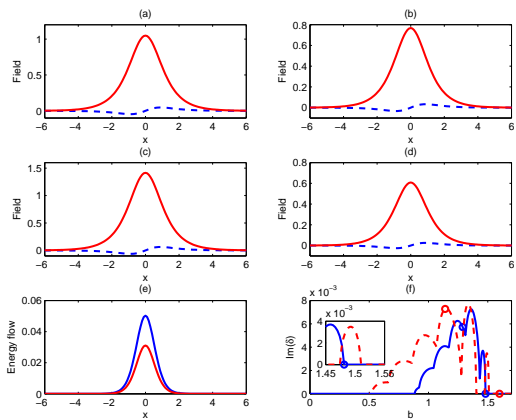}}
 \caption{\label{fig:one} (color online) (a)-(d) The real (solid red curve) and imaginary components (dash blue curve) of the soliton solutions: (a) $b=1.288$, $\alpha=2$, (b) $b=1.481$, $\alpha=2$, (c) $b=1.141$, $\alpha=5$, and (d) $b=1.599$, $\alpha=5$. (e) Transverse energy flow $S$ of solitons(blue and red curves represent the solitons in Figs.~\ref{fig:one}(b) and (d), respectively.) (f) $\text{Im}(\delta)$ versus $b$ at $\alpha=2$ (solid blue curve) and at $\alpha=5$ (dash red curve), where the points marked with circles correspond to the solitons in Figs.~\ref{fig:one}(a)-(d), and the inset depicts the curves from $b=1.45$ to $b=1.55$. The potential parameters are $V_0=3.01$ and $W_0=0.3$.}
\end{figure}

Here, we assume a Scarff II potential shown in Fig.~\ref{fig:one}(a) where $V(x)=V_{0}\text{sec}h(x)^2$
and $W(x)=W_{0}\text{sec}h(x)\text{tanh}(x)$, with $V_{0}$ and $W_{0}$
being the amplitudes of the real and imaginary part~\cite{6,7,20}. Although the PT symmetric potential has crossed the phase transition point, the solitons still exist because the amplitude of the refractive index distribution can be altered by the beam itself
through the optical nonlocal nonlinearity. The PT symmetric will remain broken if it cannot be nonlocal nonlinearly restored~\cite{6,7,9,10}.

To check the stability of the solitons
with the method of linear stability analysis, we assume $q(x)=u(x)e^{ibz}+\epsilon[F(x)e^{i\delta z}+G^*(x)e^{-i\delta^*z}]e^{ibz}$, where $\epsilon\ll1$, $F$ and $G$ are the perturbation eigenfunctions,
and $\delta$ is the growth rate of the perturbation. By linearizing Eq.~(\ref{eq:one}), we gain~\cite{6}
\begin{subequations}
\label{eq:three}
\begin{equation}
\delta F=\frac{\partial^2 F}{\partial x^2}+(V+iW)F+nF-bF+u\Delta n,\label{3a}
\end{equation}
\begin{equation}
\delta G=-\frac{\partial^2 G}{\partial x^2}+(-V+iW)G-nG+bG-u^*\Delta n.\label{3b}
\end{equation}
\end{subequations}
where $n=-\int^{+\infty}_{-\infty}g(x-\lambda)|u(\lambda)|^2d\lambda$, and
$\Delta n=-\int^{+\infty}_{-\infty}g(x-\lambda)[G(\lambda)u(\lambda)+F(\lambda)u^{*}(\lambda)]d\lambda$.
The bright spatial soltions are linearly unstable when $\delta$ has
an imaginary component, on the contrary, they are stable when $\delta$ is real.

In order to gain the solitons solutions, we numerically solve Eq.~(\ref{eq:two}) using spectral renormalization method~\cite{23}. We find a family of the simplest ground-state nonlocal bright solitons with PT symmetric potentials. To illustrate the properties of the solitons, we vary $b$, $\alpha$, $W_0$, and $V_0$. First, the typical cases of the solitons with different $b$ and $\alpha$ are shown in Fig.~\ref{fig:one}(a)-(d), where $V_0=3.01$ and $W_0=0.3$. Here the fields of  unstable and stable soliton  are respectively depicted. Evidently, the change of $b$ and $\alpha$ can influence the solutions.
\begin{figure}[htb]
\centerline{
\includegraphics[width=8cm]{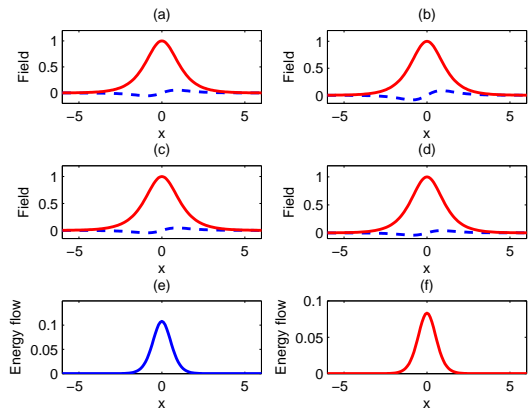}}
 \caption{\label{fig:two} (color online) (a)-(d) The real (solid red curve) and imaginary components (dash blue curve) of the soliton solutions: (a) $W_0=0.38$, $V_0=3.01$, (b) $W_0=0.60$, $V_0=3.01$, (c) $V_0=2.8$, $W_0=0.3$, and (d) $V_0=3.15$, $W_0=0.3$. (e) and (f) Transverse energy flow $S$ of solitons in Figs.~\ref{fig:one}(a) and (d), respectively. The other parameter is $\alpha=5$.}
\end{figure}
To shed more light on the properties of the stable solitons, we study the parameter $S=(i/2)(uu_x^*-u^*u_x)$, which is associated with the transverse power flow density or Poynting vector across the beam~\cite{6}. The transverse energy flow density $S$ is shown in Fig.~\ref{fig:one}(e), and the fact that S is positive everywhere implies that the energy always flows in one direction, i.e., from the gain region toward the loss region ~\cite{6}.
It is shown in Fig. 1(f) how $\text{Im}(\delta)$, the imaginary component of $\delta$, varies versus $b$. Since $\text{Im}(\delta)$ is not always equal to zero, the solitons will not always be stable. One can see that, when $\alpha$ is different, the stable regions of the solitons are different too, which means the degree of nonlocality can influence the stability of solitons.
\begin{figure}[htb]
\centerline{
\includegraphics[width=8cm]{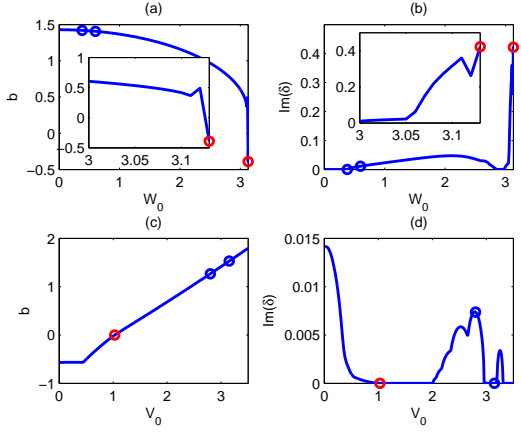}}
 \caption{\label{fig:three} (a) The propagation constant $b$ versus $W_0$ at $V_0=3.01$. (b) $\text{Im}(\delta)$ versus $W_0$ at $V_0=3.01$. (c) The propagation constant $b$ versus $V_0$ at $W_0=0.3$. (d) $\text{Im}(\delta)$ versus $V_0$ at $W_0=0.3$. Points marked with blue circles correspond to the solitons in Figs.~\ref{fig:one}(a)-(d), respectively. Points marked with red circles correspond to the threshold $W_{0upp}=3.13$ for the existence of the bright solitons where $b$ becomes negative in (a) and (b). Points marked with red circles correspond to the threshold $V_{0low}=1.03$ for the existence of the bright solitons where $b$ becomes negative in (c) and (d). The insets depict the curves from $W_0=3$ to $W_0=3.13$ in (a) and (b). The other parameter is $\alpha=5$.}
\end{figure}
\begin{figure}[htb]
\centerline{
\includegraphics[width=8cm]{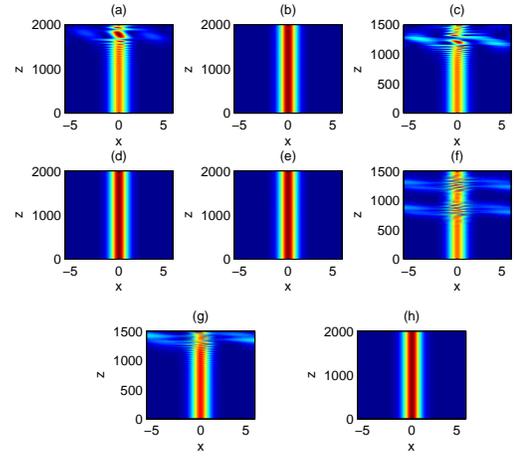}}
 \caption{\label{fig:four} (a)-(d) Simulated propagation of the solitons with $1\%$ random noise corresponding to Fig.~\ref{fig:one}(a)-(d). (e)-(h) Simulated propagation of the solitons with $1\%$ random noise corresponding to Fig.~\ref{fig:two}(a)-(d).}
\end{figure}

Next, we set $\alpha=5$ and then vary $W_0$ and $V_0$. In Figs.~\ref{fig:two}(a)-(d), the bright solitons in different conditions are shown. Figs.~\ref{fig:two}(e) and (f) depict the transverse energy flow $S$ of the solitons in Figs.~\ref{fig:one}(a) and (d), respectively. Similarly, one can find that $S$ is positive everywhere, which also means that the energy always flows in one direction. When $V_0=3.01$, we gain the relations of $b$ and $\text{Im}(\delta)$ with $W_0$ shown in Figs.~\ref{fig:three}(a) and (b), respectively. Fig.~\ref{fig:three}(a) shows that the bright solitons can exist when $0<W_0<3.13$. Thus for a fixed value of $V_0$, there exist a threshold for the imaginary amplitude $W_0$. Above this threshold ($W_{0upp}=3.13$), a phase transition occurs and the spectrum enters the complex domain~\cite{6}. Moreover, Figs.~\ref{fig:three}(c) and (d) show how $b$ and $\text{Im}(\delta)$ vary with $V_0$ when $W_0=0.3$. In Fig.~\ref{fig:three}(c), one finds that the bright solitons can exist when $V_0>1.03$, and thus, for given $W_0$, this effective potential nonlinearity will shift the PT $V_0$ threshold ($V_{0low}=1.03$) and in turn allow nonlinear eigenmodes with real eigenvalues to exist. In Fig.~\ref{fig:three}(b) and (d), $\text{Im}(\delta)$ is not always equal to zero, so the solitons are not always stable. To examine linear stability results, we further simulate the propagation of beams with random noise in the different conditions corresponding to Figs.~\ref{fig:one}(a)-(d) and Figs.~\ref{fig:two}(a)-(d), see Figs.~\ref{fig:four}.

To summary, the bright spatial solitons in nonlocal defocusing Kerr media with PT symmetric potentials are studied. The existence, stability, and propagation dynamics of such solitons are discussed in detail.


\pagebreak

\section*{Informational Fourth Page}
In this section, please provide full versions of citations to
assist reviewers and editors (OL publishes a short form of
citations) or any other information that would aid the peer-review
process.

\end{document}